\documentclass{rspublic}
\def\go{\mathrel{\raise.3ex\hbox{$>$}\mkern-14mu\lower0.6ex\hbox{$\sim$}}}
\def\lo{\mathrel{\raise.3ex\hbox{$<$}\mkern-14mu\lower0.6ex\hbox{$\sim$}}}
\def\etal{\textit{et al.\ }}
\def\eg{\textit{e.g.\ }}

\def\etc{\textit{etc.\ }}
\begin{document}
\title[Summary]{X-ray astronomy in the new Millenium.\\ 
A Summary}
\author[R. D. Blandford]{Roger Blandford}
\affiliation{Caltech, Pasadena, CA 91125, USA}
\label{firstpage}
\maketitle
\begin{abstract}{X-rays, accretion, black holes, galaxies, cosmology}
Recent X-ray observations have had a major impact on topics
ranging from protostars to cosmology. They have also 
drawn attention to important 
and general physical processes that currently limit our understanding 
of thermal and nonthermal X-ray sources. These include 
unmeasured atomic astrophysics data
(wavelengths, oscillator strengths \etc), 
basic hydromagnetic processes (\eg shock structure, reconnection), 
plasma processes (such as electron-ion equipartition and heat conduction)
and radiative transfer (in disks and accretion columns). Progress on 
these problems will probably come from integrative studies that
draw upon observations, throughout the electromagnetic spectrum,
of different classes of source.  X-ray observations are also giving a new 
perspective on astronomical subjects, like the nature of galactic nuclei 
and the evolution of stellar populations. They are contributing
to answering central cosmological questions including the 
measurement of the matter content of the universe, understanding 
its overall luminosity density, describing its chemical evolution and 
locating the first luminous objects. X-ray 
astronomy has a healthy future 
with several international space missions under construction and
in development. 
\end{abstract}
\section{Introduction}
Out of the nearly seventy octaves of electromagnetic spectrum that have been opened
up to astronomical observation, X-ray astronomers can lay claim to roughly ten
(as opposed to the single octave explored by optical astronomers!). Although no
one would pretend that all octaves are equally interesting in terms of physics, there
are some special reasons why the X-ray band is peculiarly informative. It includes
the K- and L-shell transitions of all the post-big bang elements. It is where
to find thermal emission from gas with sound speed 
$\go300$~km s$^{-1}$,
typical of the intergalactic medium, galaxies and stars. 
It lies right below $\sim m_ec^2$, which is a characteristic energy scale 
for many nonthermal processes. 

X-ray astronomy began forty years ago with the discovery of
Sco X-1 [Giaconni \etal 1962, Pounds] and was enthusiastically
developed surprisingly soon after the dawn of the space age, perhaps because
radio astronomy had, by this time, revealed a universe of sources 
with strength and properties
that were completely unexpected on the basis of optical observations. 
By the time Sco X-1 was identified, quasars and the 
microwave background had been discovered and pulsars were
soon to follow. These four discoveries ushered in modern astronomy.   

As this Discussion Meeting celebrates, X-ray astronomy has come a 
long way. It is,
arguably, ceasing to exist as a separate observational subfield of astronomy, 
so central have X-ray
observations become to the study of essentially all classes of 
cosmic sources.  The 
most improbable objects -- brown dwarfs, all types of 
protostar and the moon for 
example -- have been detected in X-rays [G\"udel]. However,
in spite of the fascination of this 
history, it is to the present and the 
future that we must turn and earlier speakers 
have taken stock of where
we are after a couple of years of full operation 
of Chandra and XMM-Newton and what the prospects are for the future.  
Even this has turned
out to be too ambitious to cover in a two day meeting and the papers presented
here (and {\em a fortiori} this brief summary) have had to be quite
selective and I shall defer to the other contributors for more representative
bibliographies.

As the only non-observer speaking at this meeting, 
I have organized my commentary
around three themes that are somewhat ``orthogonal'' to the 
preceding, source-centered talks. These are the physical processes that 
are ultimately responsible for X-ray emission, the peculiar importance of 
X-ray observations in rounding out our view of the structure and evolution
of stars and galaxies and the under-acknowledged role of X-ray astronomy 
in defining the cosmological world model to which we have been led in recent 
years and which is now starting to raise some very important questions 
concerning what actually happened in the first Gyr of the life 
of the universe. 
I conclude with a 
brief listing of the proposed next generation of X-ray observatories.
\section{Physical Processes}
The description of many high temperature and nonthermal
sources is dependent upon some poorly undestood physical processes. 
It is striking how 
often the same questions are asked of quite different physical 
environments. How
important is thermal conduction? How effective is magnetic 
reconnection in heating
plasma? And so on. The optimistic view, that links
several of the talks, is that we really only have to solve 
these problems, \eg the structure of Mach 30 shocks,
once and we
should be prepared to combine astrophysical, space physical observational with 
laboratory experiments and computational work to develop some confident 
answers.
\subsection{Spectroscopy and its Interpretation}
The gratings on XMM-Newton and Chandra have 
unprecedented spectroscopic capability,
and the ability to utilise them so effectively has benefitted from over
thirty years of hard (and largely unacknowledged) work measuring and computing
wavelengths, oscillator strengths and so on for transitions of little
terrestrial interest. Of particular importance are the Helium-like 
triplets which combine observations of permitted, intercombination 
and forbidden transitions (Gabriel \& Jordan 1969, [Kahn]) so as to provide
density and temperature diagnostics. The low densities that allow
forbidden transitions to be so important in cosmic sources 
are hard to work with experimentally. Conversely, the high radiation densities
that allow radiative ionisation equilibria to be established are only just
being achieved using powerful lasers.  

By now, most of the important wavelengths, 
oscillator strengths, collision integrals \etc for the strongest
lines have been measured,
although these measurements are still lacking for the majority of the
weaker lines that can also be observed in the brightest sources. 
Observations of
H-like and He-like transitions of the more common ions provides useful
diagnostics of the density, temperature, abundance, ionization
equilibrium and velocity in the emission regions
and we have seen here many examples 
of what can be done in a wide variety of sources including
accretion disks (stellar [Done] and AGN [Fabian]), 
stars and protostars [G\"udel], clusters of galaxies [Mushotzky] and 
polars [Cropper]. 
The power of X-ray spectroscopy is most clearly brought out 
by the detailed observations of bright supernova remnants [Canizares]. 
Here, it is
possible to use the $\sim100$~km s$^{-1}$ velocity 
resolution to make three dimensional
abundance maps of the expanding debris and 
forensic analyses of the initial explosions.
 
Unfortunately, most other sources are unresolved 
and most direct analyses of the data 
are often limited to one zone models and some very primitive radiative 
transfer. Now, it is possible, at least in principle,
to include anisotropy, inhomogeneity and peculiar geometric effects
in theoretical models of these sources. The problem is that we really have no 
clear idea of the disposition and structure of the emission region and the 
medium through which the radiation is propagating. A particularly important 
example is provided by the ongoing debate concerning the soft X-ray spectra
of Seyfert galaxies. Does the power-law, continuum source that 
is reflected by the disk arise in a local corona or at high altitude
so that it can illuminate the whole disk? Are the carbon and 
oxygen lines produced by reflection (Sako \etal 2002) or in a 
dusty, warm absorber (Lee \etal 2001) again at some distance? 
Why does there appear to be no sign of a reverberative
response in Seyfert galaxies with variable X-ray continua? Answering
these questions using a more detailed
analysis of line formation in the two cases,
is tantamount to understanding the source geometry. However,
as broad emission lines are now being reported from binary X-ray sources like
Cyg X-1 [Fabian] and J1650-500 (Miller \etal 2002), it is reasonable to
suppose a model that works for Seyferts should also work for 
black hole binaries.

A second example is provided by stellar coronae, where hot, 
coronal gas is excited by twisted, magnetic loops. (Temperatures as 
high as 40 million 
degrees are now reported associated with the Galactic Cepheid, 
YY Mon [G\"udel].) 
Although we can image similar activity in the sun, we still do not 
understand it at all
well. Again, we do not have an agreed story as to the sequence of events 
that leads to coronal lines being emitted and, consequently, 
how to convert raw line
strengths from distant stars into physical conditions in 
their coronae. The observation
of coronal activity from late M stars that are thought to be 
cool enough not to 
possess surface convection zones suggests that other, 
non-magnetic processes could be at work. Again, there is probably 
a general theory that can be inferred by combining solar and stellar
observations.    

A third case, where we probably do understand the geometry and can  
compute the emissivity and opacity, is provided by 
cyclotron line formation in accreting white dwarfs [Cropper]. Although eclipse 
observations have confirmed the expected strongly 
inverted temperature gradient,
the radiative transfer is quite subtle and a far more detailed theoretical
treatment is likely to be necessary for us to reproduce the observed spectra
as well as their polarisation and time-dependence. (Monte Carlo techniqes 
will surely continue to play a major role here and some of the experience
gained from working with Tokamak plasmas may be relevant.)
\subsection{The Plasma Impasse}
X-ray sources are, inevitably, fully ionized gases. It is therefore
unfortunate that X-ray astronomers have been resistant to learning the 
principles of plasma physics and incorporating them into their 
science, prefering instead to limit their purview to gas dynamics and 
atomic physics. This evasion can no longer
be excused. There are now several sources where progress awaits the 
answers to fundamental plasma physics questions. 
Again, I only have space for a few examples.

The first question is ``How fast do ions with some temperature $T$ 
heat cooler electrons?''.
There is a minimal and standard heating rate resulting from two body
Coulomb scattering. However, there is an abundance of wave-particle
interactions that might, for example, be excited by streams of fast 
particles. The empirical evidence comes from the observations of 
high Mach number, heliospheric and supernova remnants shock fronts, 
which appear to transmit thermal electrons with a 
temperature well below the equipartition value so that the ions do not 
quickly attain collisional ionization equilibrium [Canizares]. 
This suggests that
collective effects are not that important. This view is supported 
by observations of slowly accreting black holes which are also best interpreted
in terms of a minimal, Coulomb heating rate. The answer to this question
is of direct relevance to the debate about the efficacy of electron
heat conduction. There are really two issues. What is the mean free path
of the electrons along the magnetic field and how quickly do the 
field lines wander in response to an imposed turbulence spectrum?
As discussed below, our best laboratories are clusters of galaxies. 

A related plasma question concerns the efficiency of 
strong shocks for accelerating cosmic rays. There is a linear theory which 
appears to have some validity, again based upon heliospheric measurements
and X-ray observations of supernova remnants
[Canizares]. However, in order to model the observations,
we have to understand how the back reaction associated with the cosmic ray 
pressure moderates the acceleration and what controls the rate of both
ion and electron injection.  Theoretically, it may 
soon be possible to perform $3+3$ dimensional 
kinetic simulations with sufficient resolution to address these questions.
Observationally,
GLAST should provide measurements of the energetically-dominant  
GeV ions. As relativistic electrons are scattered by the same Alfv\'enic
turbulence that operates on the ions, the combination of $\gamma$-ray and
X-ray observations should enable us to infer the injection 
rate and perhaps determine the scaling with Mach number.
 
Another, quite controversial feature of shock fronts is their role in 
amplifying magnetic field. Theoretically, there is no very good reason
why simple gas dynamical shocks should do any more than compress 
the pre-shock magnetic field. In addition, radio observations 
of many (though not all) supernova remnants seem to show that 
the emissivity and, presumably, the magnetic field strength only increase 
in the interaction region between the shocked ejecta and the swept up 
interstellar medium. Conversely, it is possible that the hydromagnetic 
turbulence that is invoked to scatter the cosmic rays leads to an overall
increase in the rms magnetic field strength. This is highly relevant 
to the late-time evolution of $\gamma$-ray burst afterglows. 

Even greater uncertainty surrounds our understanding
of relativistic shocks, which are throught to be the primary acceleration site
for X-ray-emitting relativistic electrons
and magnetic field amplification in $\gamma$-ray 
bursts, pulsar wind nebulae and extragalactic jets.   
However, the diffusive mechanism for particle acceleration that operates
nonrelativistically is kinematically precluded. There is a promising
relativistic variant (Achterberg \etal 2001). However, 
this assumes that the cosmic rays can move far enough 
upstream from the shock to scatter off the background flow and it is 
not clear how this can happen if there is an oblique magnetic field. 
Neither is it clear how the scatterers can be generated. The magnetic
field itself is also believed to be strongly magnified at the shock front,
though no good explanation of how this happens has been found. Indeed,
the very existence of sudden, collisionless discontinuities,
as opposed to a slow sharing of momentum between two fluids,
has been questioned. X-ray observations should be especially
instructive because they permit us to resolve these putative shocks, 
say in the Crab Nebula and jets like M87, at energies where the emitting,
relativistic electrons quickly cool. The observation of 
what is presumably X-ray synchrotron radiation well away from the 
supposed strong shocks implies that relativistic electrons have 
to be accelerated {\em in situ}, rather than at strong shocks.
These observations further raise the possibility, discussed elsewhere,
(Blandford 2002)
that the observed sources including their ``shocks''
are actually relativistic, electromagnetic 
structures and are not well-described by gas dynamics.  

Another general process is magnetic reconnection which has 
been invoked, for example, in explaining the 
energisation of accretion disk coronae. However, the manner in which it 
operates remains quite controversial. Most existing discussions 
(\eg Priest \& Forbes 2000) have been 
essentially hydromagnetic except within a small region where the 
magnitude of the field
gradient becomes very large and where a scalar (and usually ``anomalous'')
resistivity is invoked.  A recent development, which has serious
implications for the topological behaviour, is that the resistivity might 
be dominated by non-dissipative, Hall terms
(Bhattarcharjee, Ma \& Wang 2001). These embellishments of 
MHD are now finding their way into numerical simulations and it will 
be interesting to see what are their implications for X-ray sources. 

In addition to numerical simulation and {\em in situ}
observation of space plasmas, it is becoming posssible to 
address some of these questions using the growing field 
of laboratory experimentation. It is 
now possible to create relativistic plasmas - both ionic and pair plasmas -
using powerful lasers, electron beams and magnetic pinches. 
Temperatures as high as 100~MeV, 
energy fluxes of $\sim100$~ZW m$^{-2}$ and $\sim1$~MT
magnetic field strengths are all attainable.
``High energy density'' investigations are likely
to become much more versatile in the coming years (\eg Takabe 2001). 
\subsection{Black Hole Accretion}
The problem of accretion onto a compact object,
specifically a black hole, is generally well-posed but has also not 
had a confident solution over the past thirty years. However, through a
combination of theoretical arguments and direct observation of accreting 
sources, it has been possible to make a lot of progress recently.
The greatest excitement has probably centred around the occasional observation 
of broad iron lines from selected, low luminosity AGN and, as reported here, 
a couple of Galactic binary X-ray sources [Fabian, Done]. (We now know that 
broad lines are not seen as commonly as once thought and that their formation
must be more complicated than envisaged in early models. The prospects
for performing useful ``reverberation mapping'' do not look good.) In 
those sources, where these features are undoubtedly seen, we can say that 
there is evidence that the second parameter that characterizes a classical 
black hole -- the spin -- is responsible for the line width. Indeed, it has 
even been argued that the role of the black hole is not just the passive one
of allowing stable orbits from which highly redshifted photons can be observed,
but is an active one in which a magnetic connection of the gas to the 
spinning hole leads to an enhanced emissivity from the innermost, and most
redshifted orbits (Wilms \etal 2002).   
 
Another very promising line of investigation 
is epitomized by the observations of Sgr 
A$^\ast$ which show that the black hole is a strikingly underluminous X-ray 
source with an apparent luminosity $\sim10^{-8}$~L$_{{\rm Edd}}$ and a
radiative efficiency of $\sim10^{-7}c^2$ relative
to the inferred mass accretion rate of $\sim10^{22}$~g s$^{-1}$. There 
have been several explanations put forward, but most of these require
that the rate of electron-ion equilibration be slow, as discussed above.
It no longer seems possible that all of the mass supplied can accrete
onto the hole and either most of the mass is lost (Blandford \& Begelman
1999) or the accretion
backs up to the Bondi radius at $\sim10^7$ gravitational radii. The whole
matter has been made more interesting through the discovery of surprisingly
rapid X-ray variability in Sgr A$^\ast$ (Baganoff \etal 2001)
and the even more remarkable suggestion
that the radio variation may be periodic (Zhao, Bowers \& Goss 2001). 
These observations open up many
more possibilities and will undoubtedly be quite constraining once the 
observational situation is clarified.

Galactic black holes provide more immediate gratification for 
observers than 
massive black holes, both on account of their larger fluxes and also 
because of their much more rapid variability timescales [Done]. 
There is now 
a fairly convincing, qualitative explanation of the low and high states.
The former arise when the luminosity is $\go0.03L_{{\rm Edd}}$ and a 
thin (or slender) disk extends down to $r_{{\rm ms}}$; the latter when
there is a central hole filled by gas that cannot cool and radiate 
efficiently and where a nonthermal spectrum is created by Comptonisation. 
It is not clear that all of this gas accretes onto the black hole.   

Many of the questions raised by these observations are issues of 
theoretical principle that are still being debated.
The approaches that will be necessary to address these questions are both
observational and theoretical. For the former, the angular resolution 
of Chandra can be put to great advantage resolving the accretion
radii in nearby, dormant galactic nuclei.  These observations are helping
us to define the physical conditions and perhaps 
to deduce the rate of gas supply to the central 
black hole. Theoretically, there are opportunities for carrying out 
3D numerical fluid dynamical and MHD/electromagnetic (including general
relativisitc) simulations of disks and outflows. 
\subsection{Nonthermal Emission}
The capability to perform arcsecond imaging at X-ray wavelengths is
revolutionizing our view of nonthermal emission. Surely, the most famous 
instance of this is the discovery of a pair of axial jets in the Crab
Nebula (Weisskopf \etal 2000)
as well as other Pulsar Wind Nebulae. This was relatively unexpected
and shows that accretion disks are not necessary for ``jet'' formation.
However, it may suggest something even more fundamental and to explain this,
I should return to one of the first models of a pulsar, the Goldreich-Julian,
axisymmetric model. Here, it was proposed that a spinning, magnetised
neutron star acts as a unipolar inductor and generates an EMF, 
${\cal E}\sim30$~PV
in the case of the Crab pulsar and that this drives a current $\sim300$~TA
around a quadrupolar circuit. (We now know, thanks to Ulysses, 
that the heliospheric electrical circuit is of this form
although the EMF and current are  only $\sim100$~MV and  $\sim1$~GA
repectively.)

Now real pulsars are, by definition, 
non-axisymmetric and the electromagnetic field just beyond the light cylinder
will contain both an AC and a DC component. The interaction between these 
two components is unclear, but it has generally been assumed that essentially
all of  the
electromagnetic Poynting flux is quickly converted into the kinetic 
energy of a plasma-dominated, outflowing wind.  In other words, the electrical
circuit is completed quite close to the pulsar. In this case, a hypersonic 
wind is created which, it has been supposed, passes through a strong 
shock front with Lorentz factor $\sim10^6$
close to the famous ``wisps'', where particle acceleration 
and nonthermal emission can occur. However, the X-ray image really shows 
no evidence for this shock front except perhaps along the poles and 
the equator.  The moving features that are seen optically also 
appear to be confined to the equatorial plane.

These observations suggest a different interpretation
(Blandford 2002), 
specifically that the AC electromagnetic
component dies away very quickly, perhaps non-dissipatively 
while the DC component persists all the way into the nebula.
In this case, the 
X-ray emission that is observed largely delineates the quadrupolar current 
flow. More specifically,
there is no strong, reverse shock and 
relatively little of the current completes 
close to the pulsar. The emission that is seen may well reflect 
MHD instability in the magnetic configuration - pinches and current sheets
are notoriously unstable - and be a manifestation of ohmic dissipation.
In this case, the observed Crab Nebula would be magnetically-, rather than
particle-dominated except, perhaps, in the emission region 
where the relativistic electron energy density might build up to 
an equipartition value. These ideas should be testable by examining the 
spectral index gradients and the polarization map. 

This viewpoint has implications for ultrarelativistic jets and gamma ray bursts
(GRB), which are now also widely acknowledged to have an electromagnetic
origin but where it is also supposed that the Poynting flux is quickly
converted to fluid form (most commonly as an optically thin pair 
plasma in the former case and as a radiation-dominated fireball in the latter)
and that the ultimate emitters are strong, relativistic shock fronts. By 
contrast, under the electromagnetic hypothesis, it is supposed that 
the energy released remains in 
an electromagnetic form all the way to the emission region and that the 
particle acceleration is a direct result of wave turbulence. There should be 
ample potential difference available for particle acceleration to take
place. In the case of a 
quasar jet, the EMF is $\sim100$~EV and the current is $\sim1$~EA. (For GRBs
the estimates are now $\sim10$~ZV and $\sim100$~EA, rather lower 
than in the past.)
%These considerations could also be relevant to unfilled supernova remnants 
%and clusters of galaxies where we are clearly observing a mixture of thermal
%and nonthermal emission and a host of fundamental questions arise concerning
%fundamental processes like cosmic ray transport, thermal conductivity, shock 
%structure, reconnection. The key point is that once we have solved these
%problems in a relatively accessible environment, we should be able to 
%export this understanding to more remote sources.
\section{Astronomical Questions}
\subsection{Nuclear Power}
High angular resolution, 
X-ray observations have, in an almost literal sense, transformed our view
of active galactic nuclei (AGN). They have amply confirmed the finding 
from optical observations, that our classification of these sources is 
strongly aspect-dependent.
The simple geometrical model of an AGN 
invokes a thick torus that will absorb UV 
continuum and emission lines and all but the hardest X-rays and $\gamma$-rays
and then
re-emit the absorbed energy in the thermal infrared. 
This description has received impressive
confirmation with ASCA, XMM and Chandra measurements of hard X-ray spectra
[Matt] which clearly exhibit the effects of absorption with 
hydrogen column densities that can approach $\sim10^{24}$~cm$^{-2}$. 
This, in turn, implies
that a significant fraction of the infrared background, as well as the
bolometric luminosity density of the young universe, be associated with 
AGN. In round numbers, the energy density associated with the
observed X-ray backgound (mostly in the energy interval $\sim20-40$~keV)
is $3\times10^{-17}$~erg cm$^{-3}$, a fraction $\sim0.003$ of the energy 
density measured in both the optical and in the far infrared backgrounds
(and $\sim10^{-4}$ of the microwave background energy density). 
In other words, if the intrinsic, integrated, UVX power of an AGN is,
on average, thirty times the HX power and this is a reasonable guess based
upon observations of local AGN, then AGN must account for $\sim10$~percent 
of the infrared background, with the remainder being presumably associated
with stars. Estimates in the literature based upon more
detailed assumptions about the mean AGN spectrum and the redshift 
distribution of the sources range from $\sim3-30$~percent.
However, this quantity may not be very well-defined
because much of the luminosity associated with galactic nuclei 
could be attributed to starbursts as opposed to accretion onto massive 
black holes [Ward].  

The characterisation of the obscuring
material as a torus is problematic, at least from a theoretical viewpoint.
The difficulty is that it is very hard to see how a thick, cold ring of 
molecular gas can be supported. Individual clouds should collide inelastically
and a torus would quickly deflate.  A more plausible 
alternative ({\em eg} Sanders \etal 1989) is that there is a 
locally thin, though strongly warped disk, where the thickness
is maintained through the marginal growth of gravitational instabilities.
However there is still confusion about the size of this disk. Combined
X-ray and SIRTF observations should greatly improve our understanding
of AGN obscuration. 

Obscuration is not the only way to produce beaming.  The X-ray study 
of relativistic jets, especially in blazars and nearby sources like 
M87 (which would probably be classified as a blazar were it pointed at us)
is becoming more 
sophisticated. However we are still not yet able to answer the quantitative 
questions about beaming fraction, the angular and radial
variation of jet Lorentz factors \etc. These questions 
will undoubtedly be a focus of future X-ray research, especially after
GLAST is launched.

The luminosity of galactic nuclei is not exclusively radiative. There
are ``superwinds'' which are driven by nuclear activity [Ward]. 
In addition, roughly ten percent of optically-selected quasars 
exhibit (X-ray quiet) broad absorption line outflows. 
Even if these flows only represent a minor
fraction of the nuclear power budget, they can still have a large impact
on the host galaxies because the momentum flux of a wind scales
inversely with the outflow speed. Not only can these outflows drive 
away much of the
interstellar medium, they may even influence the overall galaxy morphology
by, for example, inhibiting or limiting disk formation.
\subsection{Stellar Populations}
X-ray observations are also presenting us with a complementary perspective 
to that obtained using optical and infrared 
observations on how the stellar populations 
of galaxies evolve. This is because they allow us to witness the 
endpoints of stellar evolution as opposed to the beginnings. So far most
attention has been on nearby galaxies, both spirals and ellipticals [Ward].
The primary targets are young supernova remnants and binary X-ray soures.
Comparison with radio and infrared observations is particularly 
important in the former case as it allows us to derive some useful 
empirical correlations. As an example of what can be learned in this 
way, it has been reported that the star formation rate declines
less rapidly than the supernova rate [Ward].

Probably the biggest new discovery is the ultraluminous X-ray sources which
are now showing up quite regularly in all types of galaxy [Ward]. 
They are defined
by having X-ray luminosity in excess of the Eddington limit for a 
$\sim30$~M$_\odot$ black hole. They have been associated with intermediate
mass black holes, conceivably relics from the Population III
era. In this case the fuel supply is a bit of a puzzle. They must either
be short-lived binaries or single stars moving slowly through molecular 
clouds and, consequently, a major constituent of the universe, overall.
Alternatively, they could be the long-sought stellar blazars, although
here the absence of a radio emission is surprising. A third possibility
is that they are normal mass black hole binaries with super-Eddington 
luminosity which may be physically possible in the highly
clumped, radiation-dominated fluid that is expected in the innermost
regions of accretion disks orbiting $\sim10$~M$_\odot$ black holes, is 
responsible. Again, this phase would have to be quite short-lived. 
The heterogeneity of the observed properties of ultraluminous X-ray sources
suggests that more than one of these explanations could be correct. 
\section{Cosmological Issues}
\subsection{Clusters}
X-ray observations continue to be of central importance to the 
study of rich clusters of galaxies [Allen, Mushotzky]. The most basic 
information concerns the shape and depth of the gravitational
potential well. The various methods that have been used to determine
this now seem to be in fairly good agreement.
Arguably the most reliable is gravitational lensing -
strong in the core and weak in the outer parts - 
especially when there are reliable
source redshifts. As demonstrated here, this technique works well
for clusters that appear to be nearly circular and which
are dynamically relaxed.
The total mass density can then be derived from Poisson's equation. (The
smoothness of the known arcs assures us that the potential is 
also quite smooth.) It is then possible to use imaging spectroscopy to 
measure the baryon mass distribution, as roughly 85 percent of this 
is believed to be in the form of hot gas. (In practice this is carried
out by model-fitting rather than an unbiased inversion.) This procedure
may be problematic near the center of the cluster but is far safer in the 
outer parts of the cluster where most of the mass resides and where
we are most likely to sample fairly the cosmological mass distribution.
If we are prepared to trust the baryon density derived from the measured
deuterium (plus other light elements) abundance and the theory of 
big bang nucleosynthesis ($\Omega_b=0.04$), then we can 
deduce the contemporary matter
fraction of the universe and a value $\Omega_0=0.31\pm0.03$ was quoted
(assuming a Hubble constant $H_0=70$~km s$^{-1}$~Mpc$^{-1}$) [Allen].
This argument, which preceded the more publicised 
type Ia supernova determination and which
is less subject to systematic error, appears to be holding up very well. 
Repeating these measurements with larger redshift clusters,
probably in conjunction with Sunyaev-Zel'dovich measurements,  should
allow us to 
infer the expansion history of the universe, at least in principle. 
To date, this translates into an unsurprising bound on $\Omega_\Lambda$.

It is also possible to explore the thermal history of the 
intergalactic gas by comparing the measured mass -- temperature 
relation with the expectations of adiabatic simulations. It should 
not be too surprising that the agreement here is less good
[Mushotzky, Allen]. The 
entropy history of the gas which affects this determination is likely
to be quite complicated. The gas temperature, immediately after re-ionization
by the first stars and quasars is only $\sim10$~km s$^{-1}$. However as
structure grows, this gas will acquire speeds 
of several hundred km s$^{-1}$. Strong, large 
scale shock fronts are likely to form and increase the entropy of the gas.
However, as noted above, the post-shock electron temperatures will
be significantly lower than the ion temperatures 
(There is no observational evidence for strong accretion shocks
surrounding observed clusters; it appears, quite 
reasonably, that the gas is heated much earlier in the merger hierarchy.) 
The gas will also be mixed with cooler gas swept out of galaxies. It is 
very hard to compute the influence of these and other easily-imagined
effects from first principles. The dark matter, mass-velocity relation, 
which is probably best determined by gravitational lensing, is more likely
to furnish a robust measure of the growth of large scale structure.

There is an inescapable implication of the cluster gas having been heated
by strong shock fronts, whatever their provenance. This is that the hot gas
will be accompanied by cosmic ray ions. The speeds and Mach numbers of the 
shock fronts are quite similar to those in the interplanetary and 
interstellar media and we expect that the post-shock, GeV cosmic 
ray partial pressure will lie somewhere in the range 0.2-0.5 of the total
pressure. This fraction will decrease slightly as the gas 
is adiabatically compressed but 
if the gas starts to cool appreciably then cosmic rays may dominate
the pressure and inhibit further cooling. This may be part of the 
explanation of the suprising results from X-ray spectroscopy 
by XMM of a few well-studied clusters [Mushotzky]. These seem
to show that the gas starts to cool as it flows towards the central cD
galaxy and then appears to vanish -- the lines expected from gas with 
temperature below $\sim2$~keV are absent. By resisting further 
compression, the cosmic rays will make it easier to keep the gas hot.
Other factors that have
been invoked to explain the failure to observe cooling flows 
include variable metallicity, thermal conduction and supernova heating.   
There is a further
implication of having these cosmic rays present and this is that they may 
contribute to the heating and, in particular, may create $\gamma$-rays 
through pion production. The predicted $\gamma$-ray flux from nearby clusters
should be detectable by GLAST and, under extreme assumptions, could contribute
to the $\gamma$-ray background.  

X-ray observations are also providing new information on the chemical
history of the universe as we try to 
reconstruct the history of clusters of galaxies  [Mushotzky].  
C, N, O, S, Fe, Ni
have all been measured in a large sample and abundance gradients 
in a smaller number. The iron abundance has now 
stabilized at $[Fe/H]\sim0.3$ (possibly increasing 
with the size of the cluster) and is consistent with a Type Ia origin. 
The supernovae may have occured mostly in clusters with the processed 
gas being driven out in superwinds. The correlations of the relative 
abundances with the cluster properties as well as the radius are starting 
to become quite diagnostic of the evolutionary history of the cluster gas.
\subsection{X-ray Background}
Another great success for Chandra and XMM-Newton 
has been the resolution of $\sim80-90$ percent
of the X-ray background into $\sim3\times10^8$ discrete sources
[Brandt, Hasinger].  As discussed above,
most of the energy density appears to derive from black hole 
accretion in low power AGN. The redshift distribution of these sources
is controversial. On the one hand the $\sim10^8$ sources
which contribute most of the background appear to have modest 
redshifts $z<1$ [Hasinger]; on the other, the faintest sources 
are seen out to $z\sim3$ [Brandt]. In addition,
the surveys are so sensitive that nearby normal galaxies and 
luminous quasars with $z\go5$ are also found to be minor contributors
to the background.
I suspect that these statements are approximately true and 
not in contradiction.
  
The notion that most of these X-ray sources are obscured receives support from 
their association with ISO sources and bodes well for SIRTF observations. 
If, following the example above, we suppose that AGN account
for $\sim10$ percent of the infrared background then the energy produced, 
allowing for the expansion of the universe, 
corresponds to $\sim10^7$~M$_\odot c^2$
per {\em local} $L^\ast$ galaxy. If we assume that black holes
grow with $\sim10$ percent radiative efficiency, then 
we deduce a mean black hole mass per $L^\ast$ galaxy
of $\sim10^8$~M$_\odot$, roughly compatible with 
what is measured. In addition, as there are $\sim3\times10^9$ of these 
locally-specified $L^\ast$ galaxies out to $z\sim2$, where most of the 
black hole mass is grown, we conclude that a typical nucleus of one
of these galaxies is active for $\sim10^8$~yr, consistent with the 
Salpeter time.

A possible problem with this neat explanation is that if the spectrum 
below an observed energy $\sim30$~keV really does come from 
obscured sources, then they must have low redshifts as it is hard to see
how an absorption turnover could occur at a much higher energy than 
$\sim40$~keV when scattering dominates any plausible opacity. 
If this is true, then, when Swift or EXIST identify the 
sources that contribute most of the hard X-ray background,  
they should find that they have low redshifts. An alternative possibility 
is that the hard X-ray sources are mostly at $z\sim2-3$ and 
we are observing hard,
Comptonised spectra in a corona with temperature $\sim100$~keV.
Ultimately, we should like to extend this 
analysis all the way up to $\gamma$-ray energies and make smooth 
contact with studies of the $\gamma$-ray background.
\subsection{First Light}
The famous penetrating power of hard X-rays makes them excellent candidates
for probing the very early universe.  Recent observations are encouraging.
Quasars have been found with $z\sim6.5$ and the X-ray powers 
suggest that the holes exceed a billion solar masses. This is a 
constraint on theories of the growth of black holes and a timely reminder
that quasar activity must be intimately related to galaxy formation and
evolution during the $\sim0.5$~Gyr between reionisation and $z\sim6.5$
when the first quasars are seen.  

In addition, GRB have already been seen out to $z\sim4.5$,
where the redshift actually helps by allowing an observer with a fixed
response time to observe an earlier and brighter part of the evolution,
at greater emitted photon energy. There is 
optimism that Swift will identify X-ray afterglows emitted earlier than  
the light from the first observed quasars.
\section{Future Missions}
Although we look forward to many years of active service from Chandra and
XMM-Newton as well as the upcoming launches of INTEGRAL (2002), Swift (2003),
ASTRO E-2 (2005) and GLAST (2006), there are also longer range plans to 
construct more powerful telescopes like Constellation-X, 
EXIST, LOBSTER, Generation-X and XEUS. The problems discussed at this meeting
are already rewriting the scientific case for the longer term 
missions and it is hoped that this will be reflected in further
improvements in mission design and optimal use of over-subscribed 
international resources for space astronomy.  
\begin{acknowledgements}
I thank the Royal Society for support to attend this meeting
and the NSF and NASA for support under grants AST99-00866 and 5-2837.
\end{acknowledgements}

\label{lastpage}
\end{document}